\documentclass{article}

\usepackage{arxiv}
\usepackage{caption}
\usepackage{subcaption}
\usepackage[utf8]{inputenc} 
\usepackage[T1]{fontenc}    
\usepackage{url}            
\usepackage{booktabs}       
\usepackage{amsfonts}       
\usepackage{nicefrac}       
\usepackage{microtype}      
\usepackage{graphicx}
\usepackage{doi}
\usepackage[all]{xy}
\usepackage{setspace}
\usepackage{xcolor}
\usepackage{tikz}
\usetikzlibrary{cd}

\title{Direct Poisson neural networks: Learning non-symplectic mechanical systems}

\author{
Martin Šípka\\
 Mathematical Institute, Faculty of Mathematics and Physics, Charles University,\\ 
 Sokolovsk\'{a} 83, 18675 Prague, Czech Republic\\
Corresponding author: martinsipka@gmail.com\\
\And 
Michal Pavelka \\
 Mathematical Institute, Faculty of Mathematics and Physics, Charles University,\\ 
 Sokolovsk\'{a} 83, 18675 Prague, Czech Republic
\And
O{\u g}ul Esen\\
Department of Mathematics, Gebze Technical University,\\
41400 Gebze, Kocaeli, Turkey\\
\And Miroslav Grmela\\
\'{E}cole Polytechnique de Montr\'{e}al,
  C.P.6079 suc. Centre-ville,\\
 Montr\'{e}al, H3C 3A7,  Qu\'{e}bec, Canada
}

\newcommand{\shorttitle}{Direct Poisson Neural Networks}

\counterwithin*{equation}{section} 
\usepackage{amsmath,amssymb,amsfonts,latexsym,float,graphics,epsfig}

\newcommand{\XX}{\mathbf{X}}
\newcommand{\LL}{\mathbf{L}}
\newcommand{\JJ}{\mathbf{J}}
\newcommand{\UU}{\mathbf{U}}
\newcommand{\VV}{\mathbf{V}}

\newcommand{\xx}{\mathbf{x}}

\newcommand{\rr}{\mathbf{r}}

\newcommand{\MM}{\mathbf{M}}
\newcommand{\MMM}{\mathcal{M}}
\newcommand{\FFF}{\mathcal{F}}

\newcommand{\cchi}{\boldsymbol{\chi}}

\newcommand{\XXi}{\boldsymbol{\Xi}}

\begin{document}
\maketitle

\begin{abstract}
In this paper, we present neural networks learning mechanical systems that are both symplectic (for instance particle mechanics) and non-symplectic (for instance rotating rigid body). Mechanical systems have Hamiltonian evolution, which consists of two building blocks: a Poisson bracket and an energy functional. We feed a set of snapshots of a Hamiltonian system to our neural network models which then find both the two building blocks. In particular, the models distinguish between symplectic systems (with non-degenerate Poisson brackets) and non-symplectic systems (degenerate brackets). In contrast with earlier works, our approach does not assume any further a priori information about the dynamics except its Hamiltonianity, and it returns Poisson brackets that satisfy Jacobi identity. Finally, the models indicate whether a system of equations is Hamiltonian or not.
\end{abstract}

\tableofcontents

\doublespace

\section{Introduction}
The estimation of unknown parameters in physics and engineering is a standard step in many well-established methods and workflows. One usually starts with a model - a set of assumptions and equations that are considered given and then, based on the available data, estimates the exact form of the evolution equations for the system of interest. As an example, we can consider a situation where we need to estimate the mass of a star far away based on its interaction with light \cite{einstein-lens}, or when the moments of inertia of an asteroid are inferred from its rotations \cite{dermott1984}. The assumptions can be of varying complexity and the method for parameter estimation should be therefore adequately chosen.

Techniques for machine learning of dynamical systems have sparked significant interest in recent years. With the rise of neural network-related advances, several methods have been developed for capturing the behavior of dynamical systems, each with its advantages and drawbacks. A symbolic approach (for instance \cite{DiPietro2022SymplecticallySystems}) allows us to learn precise symbolic form of equations from the predefined set of allowed operations. This can be often the most efficient approach that frequently leads to an exact match between the learned and target system, but the class of captured equations is by definition limited by the algebraic operations we consider as candidates. 

Alternatively, one can learn directly the equations of motion
\begin{equation}
    \dot{\xx} = f(\xx, \theta)
\end{equation}
by learning $f$ parameterized by weights $\theta$. The function can be represented by any function approximator, in many cases by a neural network. Although this approach is very general, it does not incorporate any known physics into our procedure. There is no concept of energy of the system, no quantities are implicitly conserved, and the method thus might produce unphysical predictions. A remedy is the concept of physics-informed machine learning that constrains the neural network models so that they obey some required laws of physics \cite{kevrekidis-nature}. In particular, models of mechanical systems, which can be described by Hamiltonian mechanics, preserve several physical quantities like energy or angular momentum, as well as geometric quantities (for instance the symplectic two-form) that ensure the self-consistency of the systems. A neural-network model learning a Hamiltonian system from its trajectories that is compatible with the underlying geometry without any a priori knowledge about the system has been missing, to the best of our knowledge, and it is the main purpose of the current manuscript to introduce it. Moreover, we present several models that vary in how strictly they reproduce the underlying geometry and the degree to which these models learn a system can be used to estimate whether the system is Hamiltonian or not.

\textbf{Geometrization of a dynamical system.} A dynamical system is described by a differential equation, in particular, a mechanical system obeys Hamiltonian evolution equations. These equations are of geometric origin that is invariant with respect to changes of coordinates and which is preserved during the motion of the system. The geometry of Hamiltonian systems goes back to Sophus Lie and Henry Poincaré \cite{Lie,Poincare1901Mechanique}. Modern approaches extend to infinite-dimensional systems and provide foundations for many parts of nowadays physics \cite{AbMa78,Arnold-book,LibermannMarle,Marsden1999IntroductionSymmetry}. Formulating a mechanical system geometrically typically means finding a bracket algebra (such as symplectic, Poisson, Jacobi, Leibniz, etc.) and a generating function (such as Hamiltonian, energy, or entropy). The bracket is generally required to satisfy some algebraic conditions (Jacobi identity, Leibniz identity, etc.). However, there is no general algorithmic way to obtain the Hamiltonian formulation (even if it exists) of a given system by means of analytical calculations. So such an analysis is proper for machine learning.  

Apart from Hamiltonian mechanics, one can also include dissipation \cite{Gonzalez2018ThermodynamicallyMechanics,Moya2019LearningData,Sipka2021LearningEvolution} or extend the learning of Hamiltonian systems to control problems \cite{Zhong2019SymplecticControl}. Such an approach then, with the suitable choice of integrator ensures the conservation of physically important quantities, such as energy, momentum or angular momentum.

A reversible system is a candidate for being a Hamiltonian system.
For a reversible system, the beginning point could be to search for a symplectic manifold and a Hamiltonian function. Learning the symplectic character (if it exists) of a physical system (including particles in potential fields,  pendulums of various complexities) can be done utilizing neural networks, see, for example, \cite{Dierkes2021LearningNetworks, Greydanus2019HamiltonianNetworks}. 
The symplectic geometry exists only in even-dimensional models and due to the nondegeneracy criteria, it is very rigid. A generalization of symplectic geometry is Poisson geometry where the non-degeneracy requirement is relaxed \cite{Vaismann94,WeinsteinPoisson}. In Poisson geometry, there exists a Poisson bracket (defining an algebra on the space of functions) satisfying the Leibniz and the Jacobi identities. This generalization permits (Hamiltonian) analysis in odd dimensions. The degenerate character of Poisson geometry brings some advantages for the investigation of (total, or even super) integrability of the dynamics. 

The point of this paper is to better learn Hamiltonian mechanics. We build neural networks that encode the building blocks of Hamiltonian dynamics admitting Poisson geometry. According to the Darboux-Weinstein theorem \cite{marsden-introduction} for a Poisson manifold, there are canonical coordinates which make the Poisson bivector (determining the bracket) constant. This formulation has also geometric implications, since it determines the symplectic foliation of the Poisson manifold (see more details in the main body of this paper). Recently, Poisson neural networks (abbreviated as PNN) were proposed to learn Hamiltonian systems \cite{kevrekidis-pnn} by transforming the system in the Darboux-Weinstein coordinates. But, for many physical systems, the Poisson bracket is far from being in the canonical coordinates, and the dimension of the symplectic part of the Darboux-Weinstein Poisson bivector may be a priori unknown. For $n-$dimensional physical models, Jacobi identity, which ensures consistency of the underlying Poisson bivector, is a system of PDEs. To determine the Poisson structure, one needs to solve this system analytically, which is usually difficult, or enforce its validity while learning the Poisson bivector. 

\textbf{The goal of this work.}
The novel result of this paper is what we call a Direct Poisson Neural Network (abbreviated as DPNN) which learns the Poisson structure without assuming any particular form of the Poisson structure such as Darboux-Weinstein coordinates. Instead, DPNN learns directly in the coordinate system in which the data are provided. There are several advantages of DPNN: \textbf{(i)} We do not need to know a priori the degeneracy level of the Poisson structure (or in other terms the dimensions of the symplectic foliations of the Poisson manifold) \textbf{(ii)} it is easier to learn the Hamiltonian (energy), and \textbf{(iii)} Jacobi identity is satisfied on the data, not only in a representation accessible only through another neural network (an Invertible Neural Network in \cite{kevrekidis-pnn}). DPNN learns Poisson systems by identifying directly the Poisson bivector and the Hamiltonian as functions of the state variables. 

We actually provide three flavors of DPNNs. The least-informed flavor directly learns the Hamiltonian function and the Poisson bivector, assuming its skew-symmetry but not the Jacobi identity. Another flavor adds squares of Jacobi identity to the loss function and thus softly imposes its validity. The most geometry-informed flavor automatically satisfies Jacobi identity by building the Poisson bivector as a general solution to Jacobi identity in three dimensions. While the most geometry-informed version is typically most successful in learning Hamiltonian systems, it is restricted to three-dimensional systems, where the general solution of Jacobi identity is available. The second flavor is typically a bit less precise, and the least-informed flavor is usually the least precise, albeit still being able to learn Hamiltonian systems to a good degree of precision. 

Interestingly, when we try to learn a non-Hamiltonian model by these three flavors of DPNNs, the order of precision is reversed and the least-informed flavor becomes most precise. The order of precision of the DPNNs flavors thus indicates whether a system of equations is Hamiltonian or not.

Section \ref{sec.poisson} recalls Poisson dynamics, in particular symplectic dynamics, rigid body mechanics, Shivamoggi equations, and evolution of heavy top. Section \ref{sec.learning} introduces DPNNs and illustrates their use on learning Hamiltonian systems. Finally, Section \ref{sec.learning.NH} shows DPPNs applied on a non-Hamiltonian system (dissipative rigid body).

\section{Hamiltonian Dynamics on Poisson Geometry}\label{sec.poisson}

\subsection{General Formulation}
A Poisson bracket on a manifold $\MMM$ (physically corresponding to the state space, for instance position and momentum of the body) is a skew-symmetric bilinear algebra on the space $\FFF(\MMM)$ of smooth functions on $\MMM$ given by
\begin{equation} 
\{\bullet,\bullet\}:\FFF(\MMM)\times\FFF(\MMM)\rightarrow\FFF(\MMM). 
\end{equation}
Poisson brackets satisfy the Leibniz rule
\begin{equation}
\{F,H G\} = \{F,H\}G + H\{F,G\},
\end{equation}
and the Jacobi identity,
\begin{equation}\label{Jacobi}
\{F,\{H,G\}
+ \{H,\{G,F\}
+ \{G,\{F,H\}=0,
\end{equation}
for arbitrary functions $F$, $H$ and $G$, \cite{Marsden1999IntroductionSymmetry,Vaismann94,WeinsteinPoisson}. A manifold equipped with a Poisson bracket is called a Poisson manifold and is denoted by a two-tuple $(\MMM,\{\bullet,\bullet\})$. A function $C$ is called a Casimir function if it commutes with all other functions that is $\{F,C\}=0$ for all $F$. For instance, the magnitude of the angular momentum of a rigid body is a Casimir function.

\textbf{Hamiltonian Dynamics.}
Hamiltonian dynamics can be seen as evolution on a Poisson manifold. For a Hamiltonian function (physically energy) $H$ on $\MMM$, the Hamiltonian vector field and Hamilton's equation are
\begin{equation}
X_H(F):=\{F,H\}, \qquad \dot{\xx} = X_H(\xx)=\{\xx,H\},
\end{equation}
respectively, where $\xx \in \MMM$ is a parametrization of manifold $\MMM$. The algebraic properties of the Poisson bracket have some physical consequences. Skew-symmetry implies energy conservation,
\begin{equation}
\dot{H} = \{H,H\} =0,
\end{equation}
while the Leibniz rule ensures that the dynamics does not depend on biasing the energy by a constant. 
Referring to a Poisson bracket, one may determine the Poisson bivector field according to 
\begin{equation}
L(dF,dH):=\{F,H\}.
\end{equation}
This identification makes it possible to define a Poisson manifold by a tuple $(\MMM,L)$ consisting of a manifold and a Poisson bivector. 
In this notation, the Jacobi identity can be rewritten as $\mathcal{L}_{\XX_H} L = 0$, that is the Lie derivative of the Poisson bivector with respect to the Hamiltonian vector field is zero \cite{fecko}. In other words, the Jacobi identity expresses the self-consistency of the Hamiltonian dynamics in the sense that both the building blocks (Hamiltonian function and the bivector field) are constant along the evolution. 

Assuming a local coordinate system $\xx=(x^i)$ on $\MMM$, Poisson bivector determines Poisson matrix  $L=[L^{kl}]$ which enables us to write \cite{Weinstein-local}
\begin{equation}
L=L^{kl}\frac{\partial}{\partial x^k}\wedge \frac{\partial}{\partial x^l}.
\end{equation}
In this realization, the Poisson bracket and Hamilton's equations are written as
\begin{equation}
\{F,H\}= L^{kl}\frac{\partial F}{\partial x^k}\frac{\partial H}{\partial x^l},
\qquad 
\dot{x}^k = L^{kl}\frac{\partial H}{\partial x^l},
\end{equation}
respectively. Here, we have assumed summation over the repeated indices. 
Further, the Jacobi identity \eqref{Jacobi} turns out to be the following system of PDEs
\begin{equation}\label{eq.jac}
L^{kl}\frac{\partial L^{ij}}{\partial x^k}
+L^{ki}\frac{\partial L^{jl}}{\partial x^k}
+L^{kj}\frac{\partial L^{li}}{\partial x^k} = 0.
\end{equation}
The left-hand side of this equation is called Jacobiator, and in the case of Hamiltonian systems, it is equal to zero.
Jacobi identity \eqref{eq.jac} is a system of differential equations consisting of PDEs. In $3D$, Jacobi identity \eqref{eq.jac} is a single PDE whose general solution is known \cite{EsGhGu16}. In $4D$, Jacobi identity \eqref{eq.jac} consists of four PDEs, and there are some partial results, but for an arbitrary $n$, according to our knowledge, there is no general solution yet. We shall focus on $3D$, $4D$ and $6D$ cases in the upcoming subsections. 

\textbf{Symplectic Manifolds.}
If there is no non-constant Casimir function for a Poisson manifold, then it is also a symplectic manifold. Although we can see symplectic manifolds as examples of Poisson manifolds, it is possible to define a symplectic manifold in a direct way without referring to a Poisson manifold.  A manifold $\MMM$ is called symplectic if it is equipped with a closed non-degenerate two-form (called a symplectic two-form) $\Omega$. A two-form is called non-degenerate when
\begin{equation}
\Omega ( X,Y ) =0,  \qquad \forall X\in \mathfrak{X}(\MMM)
\end{equation}
implies $Y=0$. A two-form is closed when being in the kernel of deRham exterior derivative, $d\Omega=0$. A Hamiltonian vector field $X_{H}$ on a symplectic manifold $(\MMM,\Omega) $ is defined as
\begin{equation}\label{Hamvf}
\iota_{X_{H}}\Omega =dH,  
\end{equation}
where $\iota$ is the contraction operator (more precisely the interior derivative) \cite{fecko}. Referring to a symplectic manifold one can define a Poisson bracket  \begin{equation}\label{canPois}
\{F,H\}:=\Omega(X_F,X_H),
\end{equation}
where the Hamiltonian vector fields are defined through Equation \eqref{Hamvf}. The closedness of the symplectic two-form $\Omega$ guarantees the Jacobi identity \eqref{Jacobi}. The non-degeneracy condition of $\Omega$ puts an extra condition to the bracket \eqref{canPois} that the Casimir functions are only the constant functions, in contrast with Poisson manifolds, which may have also non-constant Casimirs. The Darboux-Weinstein coordinates show more explicitly the relationship between Poisson and symplectic manifolds in a local picture.  

\textbf{Darboux-Weinstein Coordinates.} We start with $n=(2m+k)$-dimensional Poisson manifold $\MMM$ equipped with Poisson bivector $L$. Near every point of the Poisson manifold, the Darboux-Weinstein coordinates $(x^i)=(q^a,p_b,u^\alpha)$ (here $a$ runs from $1$ to $m$, and $\alpha$ runs from $1$ to $k$) give a local form of the Poisson bivector 
\begin{equation} \label{loc-L-bf}
L=\frac{\partial }{\partial q^a}\wedge
\frac{\partial }{\partial p_a} + \frac{1}{2}\lambda
^{\alpha\beta} \frac{\partial }{\partial u^\alpha}\wedge
\frac{\partial }{\partial u^\beta}
\end{equation}
with the coefficient functions $\lambda^{\alpha\beta}$ equal zero at the origin. If $k=0$ in the local formula \eqref{loc-L-bf}, then there remains only the first term on the right-hand side and the Poisson manifold turns out to be a $2m$-dimensional symplectic manifold. Newtonian mechanics, for instance, fits this kind of realization. On the other hand, if $m=0$ in  \eqref{loc-L-bf}, then there remains only the second term which is a full-degenerate Poisson bivector. A large class of Poisson manifolds is of this form, namely Lie-Poisson structure on the dual of a Lie algebra including rigid body dynamics, Vlasov dynamics, etc. In general, Poisson bivectors have both the symplectic part as well as the fully degenerate part, for instance, the heavy top dynamics in Section \ref{sec.6D}.

When the Poisson bivector is non-degenerate, it generates a symplectic Poisson bracket, and it commutes with the canonical Poisson bivector
\begin{equation}
\LL_{can} = \begin{pmatrix} \mathbf{0} & \mathbf{I}\\ -\mathbf{I} & \mathbf{0}\end{pmatrix}
\end{equation}
in the sense that
\begin{equation}\label{eq.symp}
\LL\cdot\LL_{can} - \LL_{can}\cdot\LL = \mathbf{0}.
\end{equation}
This compatibility condition is employed later in Section \ref{sec.learning} to measure the error of DPNNs when learning symplectic Poisson bivectors.

\subsection{$3D$ Hamiltonian Dynamics}
In this subsection, we focus on three-dimensional Poisson manifolds, following \cite{EsGuGu22,GuNu93,Gu10}.
One of the important observations in $3D$ is the isomorphism between the space of vectors and the space of skew-symmetric matrices given by 
\begin{equation}
    \LL = \begin{pmatrix} 0 & -J_z & J_y\\ J_z & 0 & -J_x \\ -J_y & J_x & 0\end{pmatrix} \leftrightarrow \JJ = (J_x, J_y, J_z).
\end{equation}
This isomorphism lets us write Jacobi identity \eqref{eq.jac} as a single scalar equation 
\begin{equation}
\mathbf{J}\cdot (\nabla \times \mathbf{J})=0,  \label{jcbv}
\end{equation}%
see, for example, \cite{GuNu93,Hern2001,Hern2001b,Hern2001c}. The general solution of Jacobi identity (\ref{jcbv}) is
\begin{equation}
\mathbf{J}=\frac{1}{\phi}\nabla C  \label{Nsoln}
\end{equation}%
for arbitrary functions $\phi$ and $C$, where $C$ is a Casimir function. Hamilton's equation then takes the
particular form
\begin{equation} \label{HamEq3}
\mathbf{\dot{x}}=\mathbf{J}\times \nabla H =  \frac{1}{\phi}\nabla C \times \nabla H .
\end{equation}

Note that by changing the roles of the Hamiltonian function $H$ and the Casimir $C$ one can arrive at another Hamiltonian structure for the same system. In this case, the Poisson vector is defined as $\mathbf{J}=-(1/\phi)\nabla H$ and the Hamiltonian function is $C$. This is an example of a bi-Hamiltonian system, that manifests integrability \cite{EsGhGu16,EsGu20,Gu92,GuNu94}. A bi-Hamiltonian system admits two different but compatible Hamilton formulations. In $3D$, two Poisson vectors, say $\JJ_1$ and $\JJ_2$ are compatible if 
\begin{equation}\label{eq.cc}
    \JJ_1 \cdot (\nabla\times \JJ_2) = 
    \JJ_2 \cdot (\nabla\times \JJ_1).  
\end{equation}
This compatibility condition will be used later in Section \ref{sec.learning} to measure the error of learning Poisson bivectors in 3D by DPNNs.

\textbf{Rigid Body Dynamics.} Let us consider an example of a 3D Hamiltonian system, a freely rotating rigid body. The state variable $\MM\in\MMM$ is the angular momentum in the frame of reference co-rotating with the rigid body. The Poisson structure is 
\begin{equation}\label{eq.PB.SO3}
\{F,H\}^{(RB)}(\MM)=
- \MM \cdot \frac{\partial F}{\partial \MM}\times\frac{\partial H}{\partial \MM}, 
\end{equation}
see \cite{arnold}.
Poisson bracket \eqref{eq.PB.SO3} is degenerate because it preserves any function of the magnitude of $\MM$.
The Hamiltonian function is the energy 
\begin{equation}\label{eq.SO3.E}
H = \frac{1}{2}\left(\frac{M_x^2}{I_x}+\frac{M_y^2}{I_y}+\frac{M_z^2}{I_z}\right),
\end{equation}
where $I_x$, $I_y$ and $I_z$ are moments of inertia of the body. In this case, Hamilton's equation 
\begin{equation}\label{eq.M}
\dot{\MM} = \MM \times \frac{\partial H}{\partial \MM} 
\end{equation}
gives Euler's rigid body equation \cite{landau1}.

\subsection{$4D$ Hamiltonian Dynamics}

In four dimensions, we consider the following local coordinates $(u,x,y,z)=\left( u,\mathbf{x}%
\right)$. A skew-symmetric matrix $L$ can be identified with a couple of vectors $\mathbf{U}
=( U^{1},U^{2},U^{3}) $ and $\mathbf{V}= (
V^{1},V^{2},V^{3} )$ as
\begin{equation}\label{eq.LUV}
L=
\begin{pmatrix}
0 & -U^{1} & -U^{2} & -U^{3} \\
U^{1} & 0 & -V^{3} & V^{2} \\
U^{2} & V^{3} & 0 & -V^{1} \\
U^{3} & -V^{2} & V^{1} & 0%
\end{pmatrix}.
\end{equation}
After this identification, Jacobi identity \eqref{eq.jac} turns out to be a system of PDEs consisting of four equations \cite{EsChGuGu16}
\begin{subequations}
\begin{align}
\partial_{u}(\mathbf{U}\cdot\mathbf{V}) & =\mathbf{V}\cdot\left( \partial_{u}%
\mathbf{U}-\nabla\times\mathbf{V}\right) ,  \label{dsjk} \\[0.08in]
\nabla(\mathbf{U}\cdot\mathbf{V}) & =\mathbf{V}(\nabla\cdot\mathbf{U})-%
\mathbf{U}\times\left( \partial_{u}\mathbf{U}-\nabla\times\mathbf{V}\right) .
\label{dvjk}
\end{align}
\end{subequations}
Note that $L$ is degenerate (its determinant being zero) if and only if $\mathbf{U}\cdot\mathbf{V}=0$. So, for degenerate Poisson matrices, the Jacobi identity is satisfied if 
\begin{equation}
\nabla\cdot\mathbf{U}=0, \qquad  \partial_u\mathbf{U} -\nabla\times\mathbf{V}=\mathbf{0} .\label{jac}
\end{equation}

\textbf{Superintegrability.} Assume that a given dynamics admits two time-independent first integrals,
say $H_{1}$ and $H_{2}$. Then, when vectors $\UU$ and $\VV$ have the form 
\begin{equation}
\mathbf{U}=\nabla H_{1}\times\nabla H_{2}, \qquad  \mathbf{V}%
=\partial_{u}H_{1}\nabla H_{2}-\partial_{u}H_{2}\nabla H_{1},  \label{UandV}
\end{equation}
they constitute a Poisson bivector, and in particular Jacobi identity \eqref{jac} is satisfied. Functions $H_{1}$ and $H_{2}$ are Casimir functions. 
For a Hamiltonian function $H_{3}$, the Hamiltonian dynamics is 
\begin{subequations}
\begin{align}
\dot{u} & =-\left( \nabla H_{1}\times\nabla H_{2}\right) \cdot\nabla H_{3},
\\
\mathbf{\dot{x}} & = (\nabla H_{1}\times\nabla H_{2})\partial
_{u}H_{3}+(\nabla H_{2}\times\nabla H_{3})\partial_{u}H_{1}+(\nabla H_{3} \times\nabla
H_{1})\partial_{u}H_{2}.
\end{align}
\end{subequations}
By permuting the roles of the functions $H_1$, $H_2$, and $H_3$ (two of them are Casimirs and one of them is Hamiltonian), one arrives at two additional Poisson realizations of the dynamics. This is the case of a superintegrable (tri-Hamiltonian) formulation, \cite{GoNu01}. 

Two Poisson bivectors of a superintegrable 4D Hamiltonian system must satisfy the compatibility condition
\begin{equation}\label{eq.cc4}
\UU_1\cdot\VV_2 + \VV_1\cdot\UU_2 = 0
\end{equation}
where $\UU_{1,2}$ and $\VV_{1,2}$ are the vectors identified from formula \eqref{eq.LUV}.
We shall use this compatibility condition to measure the error of DPNNs when learning Poisson bivectors in 4D.

\textbf{Shivamoggi Equations.} 
An example of a 4D Poisson (and non-symplectic) system is the Shivamoggi equations, which arise in the context of magnetohydrodynamics, 
\begin{equation}
\dot{u}=-uy,\qquad \dot{x}=zy,\qquad \dot{y}=zx-u^{2},\qquad \dot{z}=xy,
\label{SE}
\end{equation}
see \cite{GuCh14,shivamoggi}. The first integrals of this system of equations are
\begin{equation}
H_{1}=x^{2}-z^{2},\qquad  H_{2}=z^{2}+u^{2}-y^{2},\qquad  H_{3}=u(z+x).
\label{FISE}
\end{equation}
Vectors $\mathbf{U}_{i}$ and $\mathbf{V}%
_{i}$ of Poisson matrices $N^{\left( i\right) }$ ($i=1,2,3$) for the Hamiltonian
functions $H_{1}$, $H_{2}$, and $H_{3}$ are%
\begin{align}
\mathbf{U}_{1} & =2u\left( -y,z,y\right) ,\qquad \mathbf{V}%
_{1}=2\left( u^{2},y(x+z),u^{2}-z(x+z)\right) ,  \notag \\
\mathbf{U}_{2} & =2\left( x+z\right) \left( 0,u,0\right) ,\qquad 
\mathbf{V}_{2}=2\left( x+z\right)\left( x,0,-z\right) ,  \notag \\
\mathbf{U}_{3} & =-4\left( yz,zx,xy\right) ,\qquad \mathbf{V}%
_{3}=4u\left( -x,0,z\right) ,  \label{PoiShi}
\end{align}
respectively. Note that all these three Poisson matrices are degenerate,
since $\mathbf{U}_{i}\cdot\mathbf{V}_{i}=0$ holds for all $i=1,2,3$. The
equations of motion can be written as%
\begin{equation*}
X=\vartheta N^{\left( 1\right) }\bar{\nabla}H_{1}=\vartheta N^{\left(
2\right) }\bar{\nabla}H_{2}=\vartheta N^{\left( 3\right) }\bar{\nabla}H_{3}%
\text{, \ \ \ }\vartheta=-\frac{1}{4(x+z)}
\end{equation*}
up to multiplication with a conformal factor $\vartheta$ in all three cases. 
Note that the 4D gradient is denoted by $\bar{\nabla}H=(\partial_u H,\partial_x H,\partial_y H,\partial_z H)$. 

\subsection{Semi-direct Extension to a $6D$ system}\label{sec.6D}
Six-dimensional Hamiltonian systems can again be symplectic or non-symplectic (degenerate). The former case is represented by a particle in three dimensions while the latter is for instance the heavy top dynamics. Since the evolution of a particle in 3D is canonical and thus analogical to the 2D dynamics, we shall recall only the heavy top dynamics.

A supported rotating rigid body in a uniform gravitational field is called heavy top \cite{holm-euler-poincare}. 
The mechanical state of the body is described by the position of the center of mass $\rr$ and angular momentum $\MM$. In this case, the Poisson bracket is 
\begin{equation}
\{F,G\}^{(\mathrm{HT})} (\rr,\MM) = -\MM \cdot \left(\frac{\partial F}{\partial \MM}\times \frac{\partial G}{\partial \MM}\right) - \rr \cdot \left(\frac{\partial F}{\partial \MM}\times \frac{\partial G}{\partial \rr}-\frac{\partial G}{\partial \MM}\times \frac{\partial F}{\partial \rr}\right),
\end{equation}
Even though the model is even dimensional, it is not symplectic. Two non-constant Casimir functions are $\rr^2$ and $\MM\cdot\rr$. In this case, we assume the Hamiltonian function as
\begin{equation}
H = \frac{1}{2}\left(\frac{M_x^2}{I_x}+\frac{M_y^2}{I_y}+\frac{M_z^2}{I_z}\right) + Mgl \rr\cdot \cchi,
\end{equation}
where $-g\cchi$ is the vector of gravitational acceleration. 
Hamilton's equation is then
\begin{subequations}
\begin{align}
\dot{\MM} =& \MM \times \frac{\partial H}{\partial \MM} + \rr \times \frac{\partial H}{\partial \rr} 
\\
\dot{\rr} =& -  \rr \times \frac{\partial H}{\partial \MM}. 
\end{align}
\end{subequations}


In the following sections, we apply DPNNs to the here recalled models and we show that DPNNs are capable to extract the Poisson bivector and Hamiltonian from simulated trajectories of the models. 

\section{Learning Hamiltonian systems}\label{sec.learning}
When we have a collection of snapshots of a trajectory of a Hamiltonian system, how to identify the underlying mechanics? In other words, how to learn the Poisson bivector and energy from the snapshots? Machine learning provides a robust method for such task.
It has been previously shown that machine learning can reconstruct GENERIC models \cite{Gonzalez2019LearningData,Gonzalez2018ThermodynamicallyMechanics,Moya2019LearningData}, but the Poisson bivector is typically known and symplectic. Poisson Neural Networks \cite{kevrekidis-pnn} provide a method for learning also non-symplectic mechanics, which however relies on the identification of dimension of the symplectic subdynamics in the Darboux-Weinstein coordinates and on a transformation to the coordinates. Here, we show a robust method that does not need to know the dimension of the symplectic subsystem and that satisfies Jacobi identity also in the coordinates in which the data are prescribed. Therefore, we refer to the method as Direct Poisson Neural Networks (DPNNs).

\begin{figure}[ht!]
    \centering
    \includegraphics[scale=0.6]{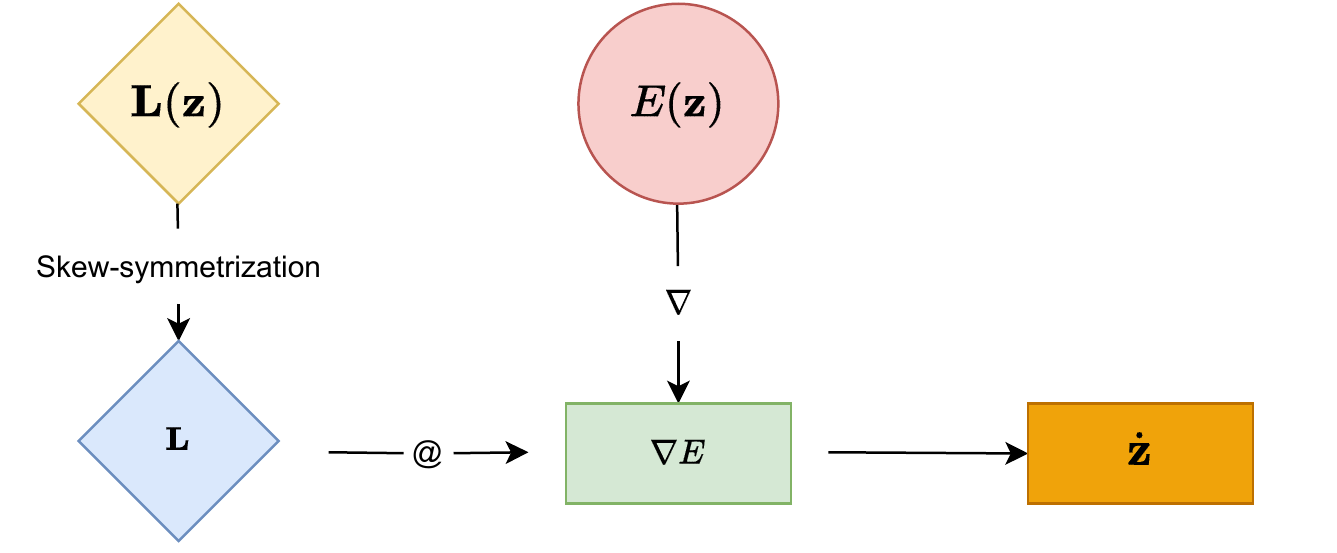}
    \caption{\label{fig.soft} Scheme SJ (Soft Jacobi) of the methods that learn both the energy and Poisson bivector.}
\end{figure}

\begin{figure}[ht!]
    \centering
    \includegraphics[scale=0.6]{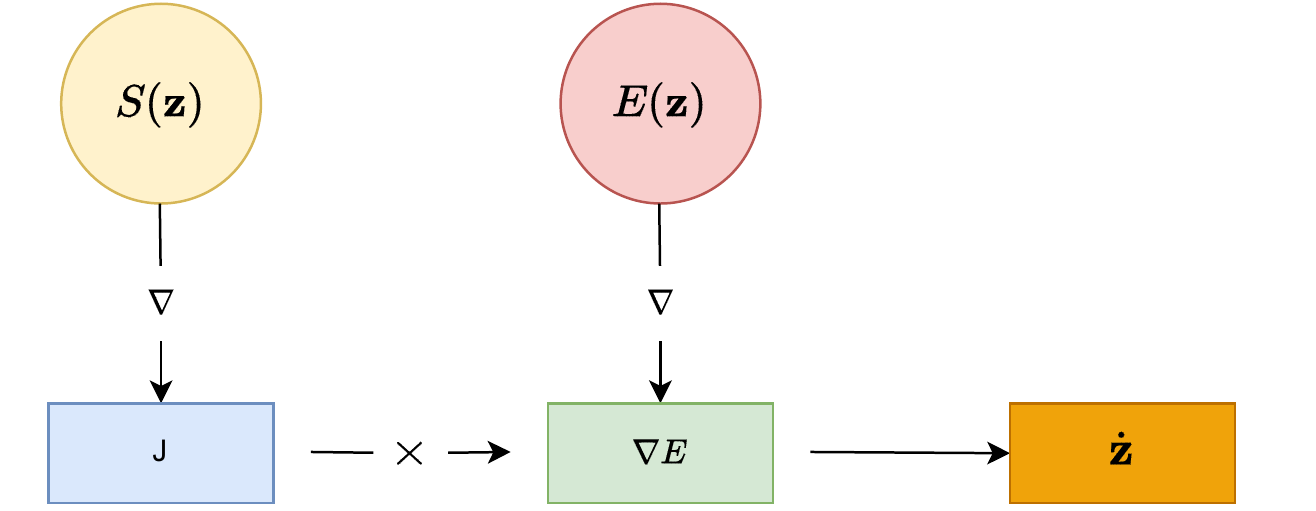}
    \caption{\label{fig.implicit} Scheme IJ (Implicit Jacobi) of the learning method implicitly enforcing Jacobi identity.}
\end{figure}

DPNNs learn Hamiltonian mechanics directly by training a model for the $\LL(\xx)$ matrix and a model for the Hamiltonian $H(\xx)$ simultaneously. The neural network that encodes $\LL(\xx)$ only learns the upper triangular part of $\LL$ and skew-symmetry is then automatically satisfied. The network has one hidden fully connected layer equipped with the softplus activation. The network that learns $H(\xx)$ has the same structure. The actual learning was implemented within the standard framework PyTorch \cite{Paszke2019PyTorch:Library}, using the Adam optimizer \cite{Kingma2014Adam:Optimization}. 
The loss function contains squares of deviation of the training data and predicted trajectories that are obtained by the implicit midpoint rule (IMR) numerically solving the exact equations (for the training data) or Hamilton's equation with the trained models for $\LL(\xx)$ and $H(\xx)$ (for the predicted data). Although such a model leads to a good match between the validation trajectories and predicted trajectories, it does not need to satisfy Jacobi identity. Therefore, we use also an alternative model where squares of the Jacobiator \eqref{eq.jac} are added to the loss function, which enforces Jacobi identity in a soft way, see Figure \ref{fig.soft}. Finally, in 3D we know the form of the Poisson bivector since we have the general solution of Jacobi identity \eqref{HamEq3}. In such a case, the neural network encoding $\LL$ can be simplified to a network learning $C(\xx)$ and Jacobi identity is automatically satisfied, see Figure \ref{fig.implicit}. 
In summary, we use three methods:
\begin{itemize}
\item \textbf{(WJ)} Training $\LL(\xx)$ and $H(\xx)$ \textit{without} the Jacobi identity.
\item \textbf{(SJ)} Training $\LL(\xx)$ and $H(\xx)$ with \textit{soft} Jacobi identity, where the $L_2$-norm of the Jacobiator \eqref{eq.jac} is a part of the loss function.
\item \textbf{(IJ)} Training $C(\xx)$ and $H(\xx)$ with \textit{implicitly} valid Jacobi identity, based on the general solution of Jacobi identity in 3D \eqref{HamEq3}.
\end{itemize}

The training itself then proceeds in the following steps:
\begin{enumerate}
\item Simulation of the training and validation data. For a randomly generated set of initial conditions, we simulate a set of trajectories by IMR. These trajectories are then split into steps and the collection of steps is split into a training set and a validation set.
\item Parameters of the neural networks WJ, SJ, and IJ are trained by back-propagation on the training data, minimizing the loss function. Then, the loss function is evaluated on the validation data to report the errors.
\item A new set of initial conditions is randomly chosen and new trajectories are generated using IMR, which gives the ground truth (GT). 
\item Trajectories with the GT initial conditions are simulated using the trained models for $\LL$ and $H$ and compared with GT. 
\end{enumerate}
In the following Sections, we illustrate this procedure for learning rigid body mechanics, a particle in 2D, Shivamoggi equations, a particle in 3D, and heavy top dynamics.

\subsection{Rigid body}
\begin{figure}[ht!]
\centering
\begin{subfigure}{.45\textwidth}
  \centering
  \includegraphics[width=1.0\linewidth]{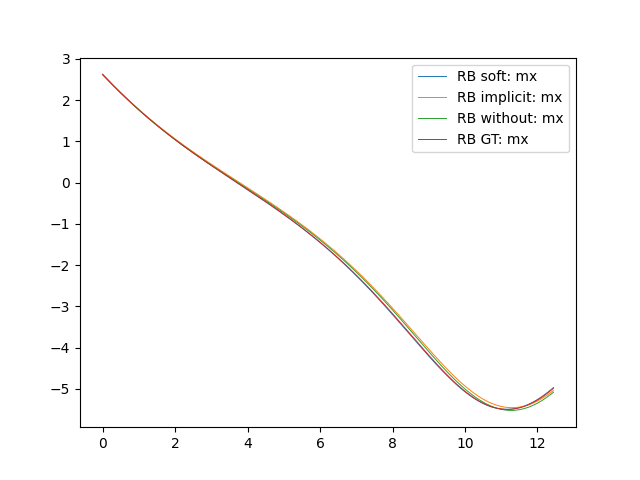}
  \caption{Comparison of an exact trajectory (GT) and trajectories obtained by integrating the learned models. All three methods fit the trajectories well.}
  \label{fig.rb.mx}
\end{subfigure}\hspace{.05\textwidth}
\begin{subfigure}{.45\textwidth}
  \centering
  \includegraphics[width=1.0\linewidth]{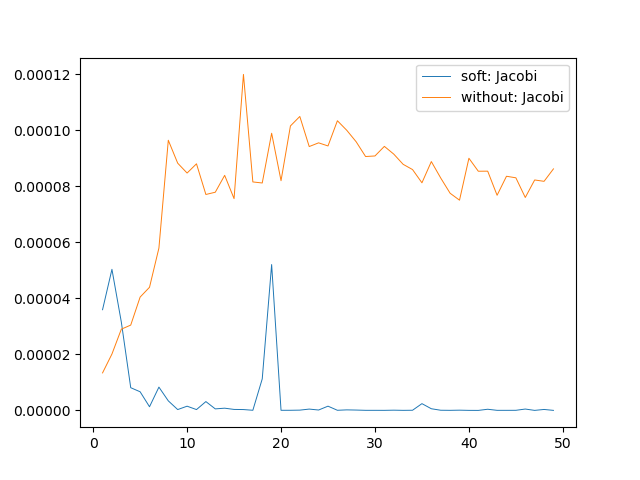}
  \caption{Error of Jacobi identity on the validation set. The error of IJ is zero by construction, the error of SJ goes to zero while the error of WJ does not.}
  \label{fig.rb.jacobi}
\end{subfigure}
    \caption{Rigid body: comparison of learned models (WJ, SJ, and IJ) with GT.}
\end{figure}

Figure \ref{fig.rb.mx} shows a sample trajectory of rigid body dynamics \eqref{eq.M} from the GT set, as well as trajectories with the same initial conditions, generated the DPNNs. The training was carried out on 200 trajectories while GT consisted of 400 trajectories. Errors of the three learning methods (WJ, SJ, and IJ) are shown in Table \ref{tab.results}. All three methods were capable to learn the dynamics well. Figure \ref{fig.rb.jacobi} shows the norm of the Jacobiator evaluated on the validation set. Jacobiator is zero for IJ and small in SJ, while it does not go to zero in WJ. Therefore, IJ and SJ satisfy Jacobi identity while WJ does not.

\begin{figure}[ht!]
\centering
  \includegraphics[width=1.0\linewidth]{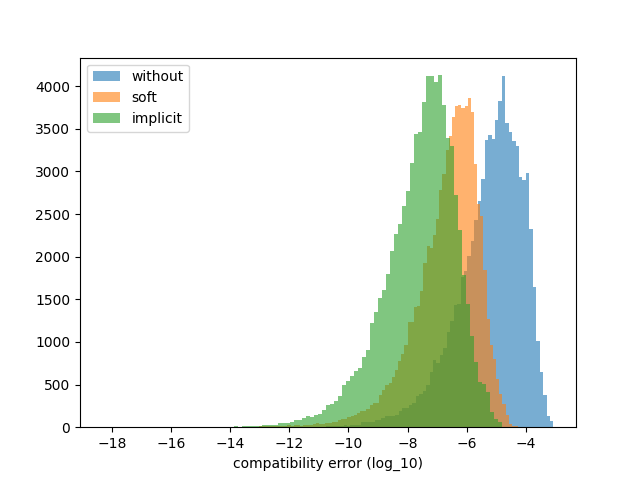}
  \caption{Rigid body: Compatibility errors for RB evaluated as $\log_{10}$ of squares of Equation \eqref{eq.cc}. The distribution of errors is approximately log-normal. The Compatibility error of the IJ method is the lowest, followed by SJ and WJ.}
  \label{fig.rb.comp}
\end{figure}

\begin{figure}[ht!]
  \centering
  \includegraphics[width=1.0\linewidth]{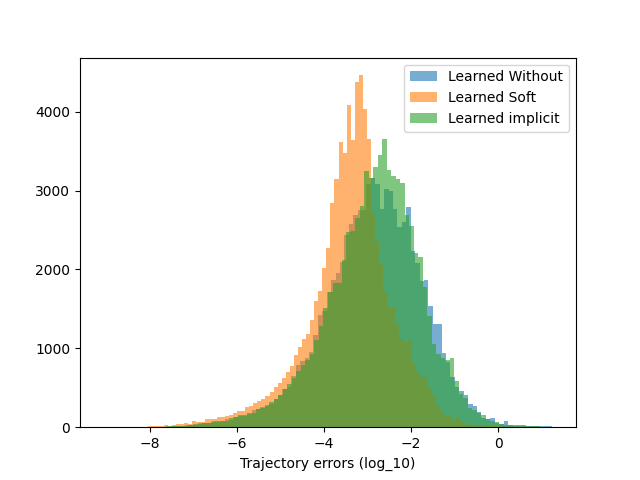}
  \caption{Rigid body: Distribution of $\log_{10}$ of squares of errors in $\MM$.}
  \label{fig.rb.m.errors}
\end{figure}

Figure \ref{fig.rb.comp} shows the compatibility error of learning the Poisson bivector \eqref{eq.cc}. All three methods learn the Poisson bivector well, but IJ is the most precise, followed by SJ and WJ. 
Finally, Figure \ref{fig.rb.m.errors} shows errors in learning the trajectories $\MM(t)$. All three methods learn the trajectories well, but in this case, the SJ method works slightly better.

Table \ref{tab.results} shows the medians of the errors for all the models and applied learning methods. N/A indicates quantities that are not to be conserved. Error $\Delta\MM$ is calculated as the median of square deviation of $\MM^2$ over all time steps. Error $\Delta\rr$, $\Delta \MM\cdot\rr$, $\Delta \MM^2$, and $\Delta \rr^2$ are calculated analogically. Error $\Delta\LL$ in the RB case is calculated as $\log_{10}$ of the $L^2$ norm of the compatibility condition \eqref{eq.cc}, calculated for the learned $\JJ$ divided by its trace and multiplied by factor 1000 (using the exact $\JJ$ when generating the GT). In the P2D and P3D cases, where the Poisson bivector is symplectic, the error is calculated as $Log_{10}$ of squares of the symplecticity condition \eqref{eq.symp}. In the case of Shivamoggi equations, the $\Delta \LL$ error is the $Log_{10}$ of the squared superintegrable compatibility condition \eqref{eq.cc4}. $\Delta \det\LL$ errors are medians of squares of learned $\det\LL$, and in the Shivamoggi and the heavy top cases, the values are logarithmic, since determinants are supposed to be zero in those cases.
\begin{figure}[ht!]
    \centering
    \begin{tabular}{|c|c|c|c|c|c|c|c|c|c|}
        \hline
        Mod. & Met. & $\Delta\MM$ & $\Delta\rr$ & $\Delta\LL$ & $\Delta\MM^2$ & $\Delta(\rr\times\MM)$ & $\Delta\rr^2$ & $\Delta(\MM\cdot\rr)$ & $\det \LL$\\\hline\hline
        RB & WJ & 1.5E-03 & N/A     & -5.0 &5.2E-04 & N/A     & N/A & N/A & 0\\
           & SJ & 4.6E-04 & N/A     & -6.5 &3.9E-04 & N/A     & N/A & N/A & 0\\
           & IJ & 1.7E-03 & N/A     & -7.4 &4.8E-04 & N/A     & N/A & N/A & 0\\\hline
       P2D & WJ & 3.2E-03 & 3.2E-03 & -1.8 & N/A    & 9.7E-03 & N/A & N/A & 0.83\\
           & SJ & 2.9E-03 & 2.9E-03 & -1.8 & N/A    & 8.6E-03 & N/A & N/A & 0.93\\\hline
        Sh & WJ & N/A     & 3.4E-04 & -2.3 & N/A    & N/A     & N/A & N/A & -2.1\\
           & SJ & N/A     & 3.0E-04 & -3.3 & N/A    & N/A     & N/A & N/A & -3.4\\\hline
       P3D & WJ & 1.9E-02 & 1.9E-02 & -1.6 & N/A    & 3.7E-02 & N/A & N/A & 0.91 \\
           & SJ & 1.0E-02 & 9.1E-03 & -1.9 & N/A    & 3.5E-02 & N/A & N/A & 0.70 \\\hline
        HT & WJ & 3.1E-02 & 9.9E-03 & N/A  & N/A    & N/A   &1.9E-03&2.8E-03& -2.5\\
           & SJ & 1.6E-02 & 4.6E-03 & N/A  & N/A    & N/A   &1.3E-03&1.9E-03& -2.6\\\hline
    \end{tabular}
    \caption{\label{tab.results}Summary of the learning errors for a rigid body (RB), particle in two dimensions (P2D), Shivamoggi equations (Sh), particle in three dimensions (P3D), and heavy top (HT).} 
\end{figure}
\clearpage

\subsection{Particle in 2D}
\begin{figure}[ht!]
\centering
  \centering
  \includegraphics[width=0.8\linewidth]{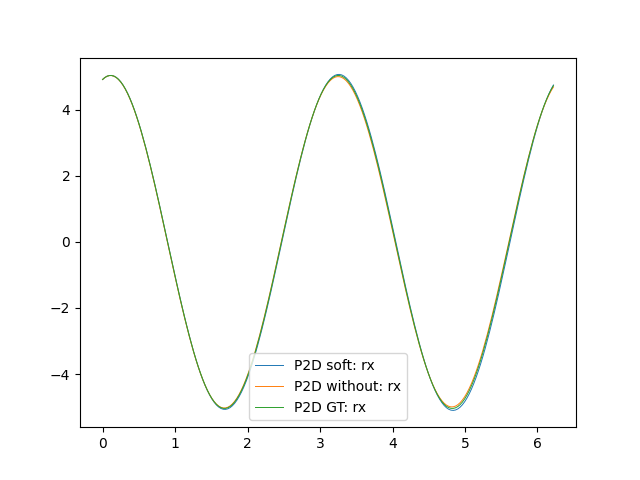}
    \caption{P2D: A sample trajectory. Both SJ and WJ learn the dynamics of a particle in two dimensions well.}
  \label{fig.p2d.rx}
\end{figure}

\begin{figure}[ht!]
  \centering
  \includegraphics[width=1.0\linewidth]{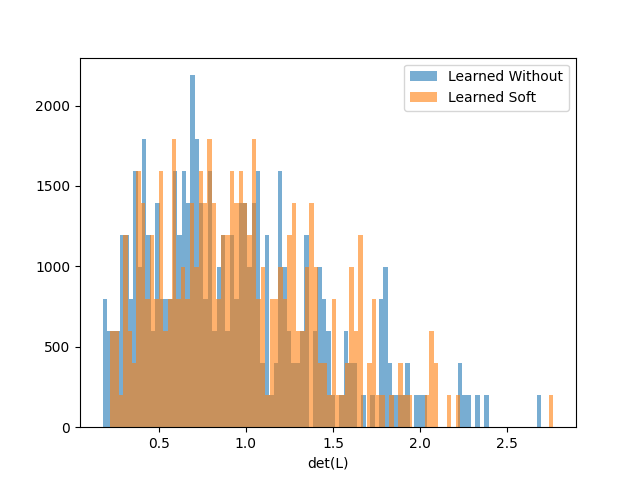}
    \caption{P2D: Learned $\det(\LL)$.}
    \label{fig.p2d.det}
\end{figure}

A particle moving in a 2D potential field represents a four-dimensional symplectic system. The simulated trajectories were learned by WJ and SJ methods. No implicit IJ method was used because no general solution of Jacobi identity in 4D is available that would work for both the degenerate and symplectic Poisson bivectors. Results of the learning are in Table \ref{tab.results}, and both WJ and SJ learn the dynamics comparably well. Figure \ref{fig.p2d.rx} shows a sample trajectory, and Figure \ref{fig.p2d.det} shows the distribution of learned $\det(\LL)$. The median determinant (after a normalization such that the determinant is equal to $1.0$ in GT), was close to this value for both SJ and WJ, indicating a symplectic system.

\subsection{Shivamoggi equations}
\begin{figure}[ht!]
\centering
  \centering
  \includegraphics[width=0.8\linewidth]{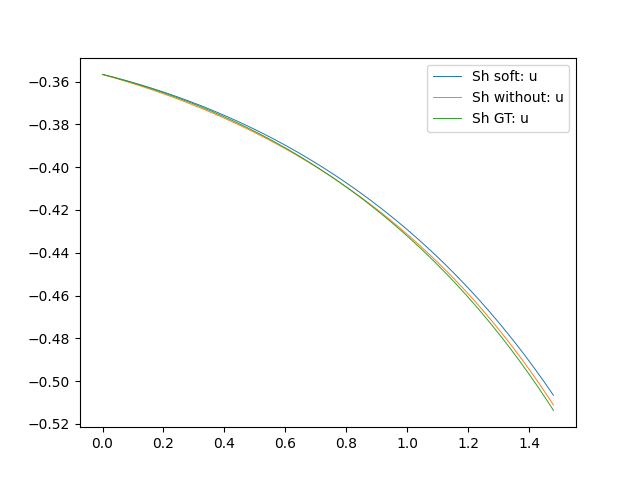}
    \caption{Shivamoggi: A sample trajectory, component $u$.}
  \label{fig.sh.u}
\end{figure}

\begin{figure}[ht!]
  \centering
  \includegraphics[width=0.8\linewidth]{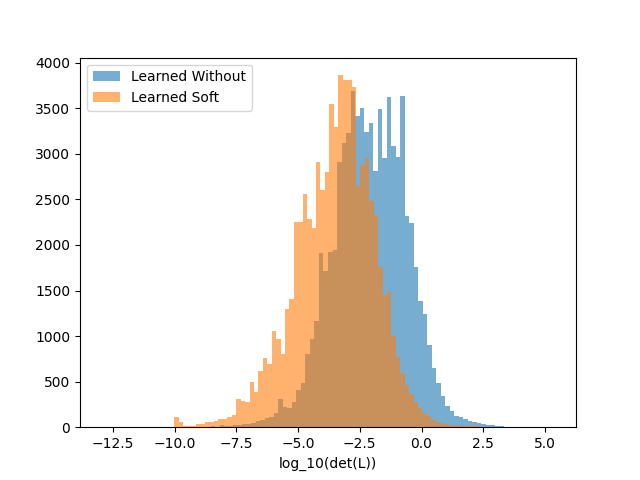}
    \caption{Shivamoggi: Learned $\log_{10}(\det(\LL))$.}
    \label{fig.Sh.det}
\end{figure}

Shivamoggi equations \eqref{SE} represent a 4D Hamiltonian system that is not symplectic, and thus has a degenerate Poisson bivector. The equations were solved within the range of parameters $u\in [-0.5, 0.5]$, $x\in [-0.5,0,5]$, $y\in [-0.1, 0.1]$, $z\in [-0.5,0.5]$. It was necessary to constraint the range of $\rr=(u,x,y,z)$ because for instance when $u=0$, the solutions explode \cite{GuCh14}. Figure \ref{fig.sh.u} shows the $u$-component over a sample trajectory. 
Figure \ref{fig.Sh.det} shows the distribution of $\log_{10}(\det(\LL))$, indicating that the system is indeed degenerate.

In comparison with determinants of $\LL$ learned in the symplectic case of a two-dimensional particle (P2D), see Table \ref{tab.results}, the learned determinants are quite low in the Shivamoggi case (after the same normalization as in the P2D case). Therefore, DPNNs are able to distinguish between symplectic and non-symplectic Hamiltonian systems.
\clearpage

\subsection{Particle in 3D}
\begin{figure}[ht!]
\centering
  \centering
  \includegraphics[width=0.8\linewidth]{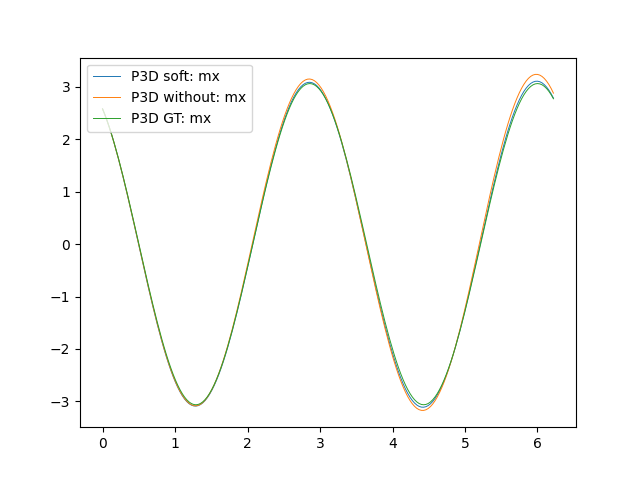}
    \caption{P3D: Comparison of momentum $\MM(t)$ on an exact trajectory (GT) and trajectories obtained by integrating the learned models (without Jacobi and with soft Jacobi) in the case of a 3D harmonic oscillator.}
  \label{fig.p3d.mx}
\end{figure}

Figure \ref{fig.p3d.mx} shows momentum during a sample trajectory of a particle in 3D space taken from the GT set, as well as trajectories with the same initial conditions obtained by DPNNs (WJ and SJ flavors). The training and validation were done in two sets of trajectories (with 200 and 400 trajectories, respectively). 
Table \ref{tab.results} contains numerical values of the learning errors. Both WJ and SJ learn the Poisson bivector as well as the trajectories (and thus also the energy) well. The median determinant is close to unity, which indicates a symplectic system.

\subsection{Heavy top}
\begin{figure}[ht!]
\centering
\begin{subfigure}{.47\textwidth}
  \centering
  \includegraphics[width=1.0\linewidth]{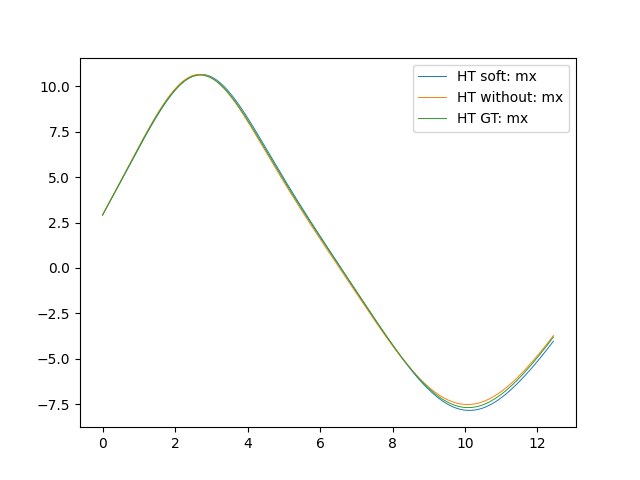}
    \caption{Angular momentum $\MM(t)$ during a sample trajectory.}
  \label{fig.ht.mx}
\end{subfigure}\hspace{.04\textwidth}
\begin{subfigure}{.47\textwidth}
  \centering
  \includegraphics[width=1.0\linewidth]{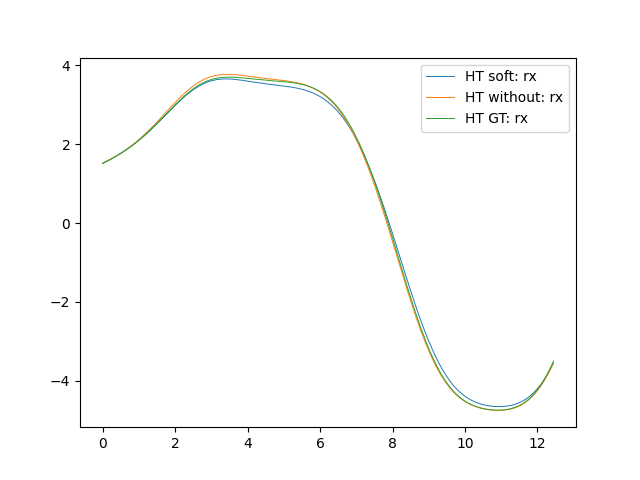}
    \caption{Vector $\rr(t)$ during the same trajectory.}
    \label{fig.ht.rx}
\end{subfigure}
    \caption{Heavy top: Comparison on an exact trajectory (GT) and trajectories obtained by integrating the learned models (without Jacobi and with soft Jacobi) in case of the heavy top.}
\end{figure}

Figures \ref{fig.ht.mx} and \ref{fig.ht.rx} show a sample trajectory of a heavy top from the GT set and trajectories with the same initial conditions obtained by DPNNs. The training and validation were done in two sets of trajectories (with 300 and 400 trajectories, respectively). Numerical values of the learning errors can be found in Table \ref{tab.results}. For instance, the $\LL$ matrix is close to being singular, indicating a non-symplectic system, but SJ learns slightly better than WJ. Similarly, as in the four-dimensional case, DPNNs distinguish between symplectic (P3D) and non-symplectic cases (HT).

\section{Learning non-Hamiltonian systems}\label{sec.learning.NH}
Let us now try to apply the WJ, SJ, and IJ methods, that are developed for learning purely Hamiltonian systems, to a non-Hamiltonina system, specifically a dissipative rigid body. A way to formulate the dissipative evolution of a rigid body is called the energetic Ehrenfest regularization \cite{Pavelka2019EhrenfestSystems}, where the Hamiltonian evolution of a rigid body is supplemented with dissipative terms that keep the magnitude of angular momentum constant while dissipating the energy. The evolution equations are
\begin{equation}\label{eq.RBdis}
\dot{\MM} = \MM\times E_\MM - \frac{\tau}{2}\XXi\cdot E_\MM
\end{equation}
where $\tau$ is a positive dissipation parameter and where $\XXi = \LL^T d^2 E \LL$ is a positive symmetric definite matrix (assuming that energy be positive definite) constructed from the Poisson bivector of the rigid body $L^{ij} = -\epsilon^{ijk}M_k$ and energy $E(\MM)$. These equations satisfy that $\dot{\MM}^2 = 0$ while $\dot{E}\leq 0$, and their solutions converge to pure rotations around the principal axis of the rigid body (an axis with the highest moment of inertia), which is the physically relevant solution.

\begin{figure}
\centering
\includegraphics[scale=0.5]{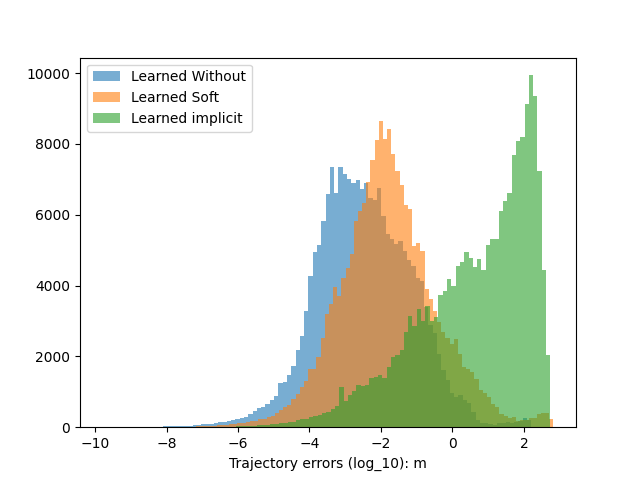}
\caption{Distribution of errors in the angular momentum $\MM$ when learning dissipative rigid-body dynamics \eqref{eq.RBdis} by methods assuming purely Hamiltonian systems (WJ, SJ, and IJ). WJ method is the most robust, capable to learn also the dissipative system relatively well (although worse than in the purely Hamiltonian case). The SJ method, which moreover softly imposes Jacobi identity, is less capable to learn the dissipative system. The IJ method, which has the best performance in purely Hamiltonian systems, see Table \ref{tab.results}, has the worst learning capability in the dissipative case. }
    \label{fig.RBdis}
\end{figure}

Results of learning trajectories generated by solving Equations \eqref{eq.RBdis} are shown in Figure \ref{fig.RBdis}. All the methods (WJ, SJ, and IJ) are capable to learn the trajectories to some extent, but WJ is the most successful, followed by SJ and IJ. As SJ, and especially IJ, use deeper properties of Hamiltonian systems (soft and exact validity of Jacobi identity), they are less robust in the case of non-Hamiltonian systems. 

Figure \ref{fig.RBdis} can be actually seen as an indication of non-Hamiltonianity of Equations \eqref{eq.RBdis}. Systems where IJ learns best, followed by SJ and WJ much more likely to be Hamiltonian, in contrast with non-Hamiltonian systems where WJ learns best, followed by SJ and IJ. In other words, DPNNs can distinguish between Hamiltonian and non-Hamiltonian systems by the order in which the flavors of DPNNs perform.

\clearpage

\section{Conclusion}
This paper proposes a machine learning method for learning Hamiltonian systems from data. Direct Poisson Neural Networks (DPNN) learn directly the Poisson bivector and Hamiltonian of the mechanical systems with no further assumptions about the structure of the systems. In particular, DPNN can distinguish between symplectic and non-symplectic systems by measuring the determinant of the learned Poisson bivector. 

DPNNs come in three flavors: (i) without Jacobi identity (WJ), (ii) with softly imposed Jacobi identity (SJ), and with implicitly valid Jacobi identity (IJ). Although all the methods are capable to learn the dynamics, only SJ and IJ satisfy also the Jacobi identity. Typical behavior is that IJ learns Hamiltonian models most precisely, see Table \ref{tab.results}, followed by SJ and WJ. 

When the three flavors of DPNNs are applied to learn a non-Hamiltonian system, it is expected that the order of precision gets reversed, making WJ the most precise, followed by SJ and IJ. This reversed order of precision can be used as an indicator that distinguishes between Hamiltonian and non-Hamiltonian systems. 

In future, we would like to extend DPNNs to systems with dissipation prescribed by gradient dynamics.

\section*{Acknowledgment}
We are grateful to E. Cueto, F. Chinesta, and B. Moya for inspiring discussions about the purpose of Jacobi identity in the learning of physical systems.
MP and MŠ were supported by the Czech Grant Agency, grant number 23-05736S.

\bibliographystyle{unsrt}
\bibliography{classic_references.bib}

\end{document}